\documentclass[twocolumn,showpacs,preprintnumbers,amsmath,amssymb,showkeys]{revtex4}

\usepackage{graphicx}
\usepackage{dcolumn}
\usepackage{bm}
\usepackage{epsfig}
\usepackage{amsmath}

\begin{document}
\title{Scroll Waves Pinned to Moving Heterogeneities}
\author{Hua Ke}
\author{Zhihui Zhang}
\author{Oliver Steinbock}
\affiliation{Florida State University, Department of Chemistry and Biochemistry, Tallahassee, FL 32306-4390}

\date{\today}


\begin{abstract}
Three-dimensional excitable systems can selforganize vortex patterns that rotate around one-dimensional phase singularities called filaments. In experiments with the Belousov-Zhabotinsky reaction and numerical simulations, we pin these scroll waves to moving heterogeneities and demonstrate the controlled repositioning of their rotation centers. If the pinning site extends only along a portion of the filament, the phase singularity is stretched out along the trajectory of the heterogeneity which effectively writes the singularity into the system. Its trailing end point follows the heterogeneity with a lower velocity. This velocity, its dependence on the placement of the anchor, and the shape of the filament are explained by a curvature flow model.
\end{abstract}
\maketitle

Processes far from equilibrium can create complex patterns that are difficult to predict from their atomistic or local dynamics. This emergence of spatial complexity often results from comparably simple transport processes. A classic example are reaction-diffusion media which generate dissipative structures such as stationary Turing patterns, traveling waves, and spatio-temporal chaos \cite{epstein14,rondelez13,cross93}. These structures are universal in the sense that they are observed across a wide range of physical, chemical, and biological experiments. Specifically, rotating spiral waves of excitation are observed in systems as diverse as active galaxies \cite{galaxies}, catalytic reactions \cite{catalysis}, and bee colonies \cite{bees}. In addition, they can orchestrate important biological functions such as the timing of contraction waves during child birth \cite{birth} or induce life-threatening conditions such as cardiac arrhythmias \cite{cardiac}.

While spiral waves have been studied intensively over the past decades, their three-dimensional counter-parts have attracted less attention. These scroll waves rotate around one-dimensional phase singularities called filaments. In general, these space curves are not static but move according to their local curvature $\kappa$ and difference in rotation phase (``twist'') \cite{biktashev94,keenertyson,pertsov97}. In simple cases, this motion obeys

\begin{equation}
    \frac{\textrm{d}\textbf{s}}{\textrm{d}t} = \alpha \kappa \hat{\textbf{N}},
\end{equation}

\noindent where $\textbf{s}$, $\hat{\textbf{N}}$ and $\alpha$ denote the filament position, its unit normal vector, and a system-specific line tension, respectively. Negative values of $\alpha$ can induce a turbulent motion of the filament \cite{biktashev94,alonso03}, whereas positive values cause curve shrinking dynamics for which filament loops annihilate and filaments connecting external surfaces converge to straight lines.

Recent studies show that filaments can attach to inactive heterogeneities \cite{pertsov00,jimenez09}. Most experiments on this type of vortex pinning employ the Belousov-Zhabotinsky (BZ) reaction \cite{mikhailov06} which is an important model of excitable and oscillatory reaction-diffusion media. Scroll waves, however, exist also in biological systems such as the human heart \cite{cardiac} for which pinning could occur at anatomical features (e.g. blood vessel and papillary muscle insertion points) as well as infarction-induced remodeled myocardium. Regardless of the specific system, pinning of scroll waves implies wave rotation around the heterogeneity whereas simple filament termination is observed at heterogeneities much larger than the free rotation orbit \cite{nakouzi14}. Pinning is subject to topological constraints, alters the rotation frequency, reshapes the global wave field, and potentially induces twist \cite{vinson93,nakouzi14,jimenez09,dutta11}. Recent studies have also shown that scroll waves self-wrap around thin cylindrical heterogeneities \cite{jimenez12} and unpin due to advective perturbations such as external electric fields \cite{jimenez13}.

In this Letter, we report the pinning of scroll waves to moving heterogeneities and show that a partially pinned filament stretches out along the trajectory of the anchor. The tail end of the filament does not remain stationary but follows the heterogeneity at a speed that is independent of the anchor speed. Its velocity and shape depend on geometric aspects and the curvature flow dynamics of the homogeneous system. These experimental and numerical results open up interesting possibilities for the study of excitable systems with dynamic heterogeneities.

\begin{figure}
\centering
\includegraphics[width=0.7\linewidth]{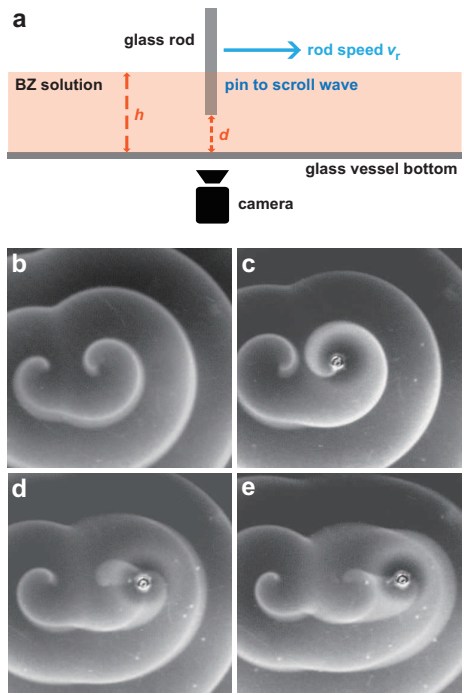}
\caption{\label{fig1} (color online) (a) Schematic drawing of the experimental set-up. (b)-(e) Image sequence of two scroll waves. The right vortex is pinned to a rightward moving glass rod. Time between subsequent frames: 20, 48, and 87~min. Field of view: 2.3~cm~$\times$ 2.3~cm.}
\end{figure}


Our experiments use a thick layer of BZ solution in a cylindrical glass vessel (diameter 5.6~cm). The system has a free solution-air interface and its viscosity is increased by addition of xanthan gum (0.4~\% w/v) and agar (0.05~\% w/v). The initial concentrations of the reactants are: [NaBrO$_3$]~= 62~mmol/L, [H$_2$SO$_4$]~= 175~mmol/L, [malonic acid]~= 48~mmol/L, and [Fe(phen)$_3$SO$_4$]~= 37.5~mmol/L. Details regarding the chemical preparation and viscosity measurements have been published in \cite{ke14}. All experiments are carried out at room temperature. We use a monochrome video camera equipped with a dichroic blue filter to monitor the chemical wave patterns. The heterogeneity is a vertical glass rod (diameter 1.1~mm) attached to a motor-driven linear actuator. The rod is submerged into the solution from the top down to create a constant gap of depth $d$ between the bottom of the rod and the surface of the container base [Fig.~1(a)]. In our experiments, we vary the value of $d$ between 0.2 and 0.75~cm while keeping $h$ constant at 1.1$\pm$0.1~cm.


Figures~1(b-e) show an image sequence of a pair of counter-rotating scroll waves in a thick layer of the BZ solution. Their rotation period is 320$\pm$30~s and their wavelength is about 0.5~cm. Initially both filaments are linear and oriented parallel to the optical axis of our set-up. The associated wave fields are untwisted. Accordingly the three-dimensional vortices are detected as simple spiral-shaped patterns (b). We then pin the right vortex to a glass rod which appears as a small disk-shaped region in (c). We emphasize that the rod does not touch the bottom of the reaction vessel but by choice, generates a gap $d$ of 0.45~mm. After five rotation periods, we begin to translate this anchor rightwards at a constant speed of $v_r$~= 0.1~mm/min (d). In response, the pinned scroll wave loses its initial, pseudo-two-dimensional character and a diffuse, bright (excited) region is formed in the wake of the anchor. We continue to observe wave rotation around the moving rod (see movie in \cite{SuppMat}) but also detect a trailing spiral-shaped feature (e). Notice that the unpinned vortex on the left is essentially unaffected by these processes.

We interpret the observed deformation of the pinned scroll wave as the result of an increasingly deformed filament. While its top portion is anchored to the moving glass rod, its unpinned connection to the base of the reaction vessel becomes stretched out along the trajectory of the rod. This stretching process is governed by i) the topological requirement of a continuous filament connection between the glass rod and the lower system boundary and ii) the flux-related requirement that filaments at Neumann boundaries must terminate in normal direction to the boundary. Accordingly, the pattern in Fig.~1(e) can be understood as a pinned (and probably twisted) scroll wave in the top portion of the system, a more horizontally oriented filament left of the anchor, and a down-curving filament terminus near the lower system boundary. The latter two regions account for the broad and diffuse feature behind the rod and its spiral-shaped termination.

\begin{figure}
\centering
\includegraphics[width=0.9\linewidth]{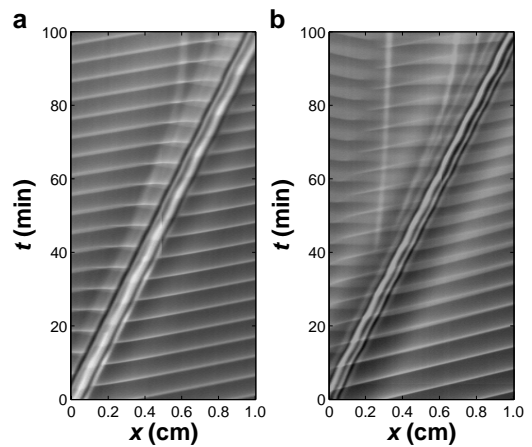}
\caption{\label{fig1} Space-time plots of scroll waves pinned to a moving glass rod. The intensity profiles are obtained along the trajectory of the rod. The experiments in (a) and (b) differ only in the gap height underneath the rod, which equals $d \approx$~0 and 0.45~mm, respectively.}
\end{figure}

The dynamics of scroll waves pinned to moving heterogeneities are further analyzed in Fig.~2. Both space-time plots are constructed from intensity profiles along the trajectory of the rod but describe an experiment with a negligibly small gap underneath the rod in (a) and the experiment shown in Fig.~1(b)-(e) for which $d$~= 0.45~mm. The moving rod itself generates the bright, diagonal band that connects the lower left to the upper right corner of the plots. The thinner bright bands result from excitation waves of the pinned scroll wave. Notice the V-shaped features left of the rod in (b) that are absent in (a). These features are caused by the alternating emission of left- and rightward moving pulses and are hence evidence for a rotating vortex. Accordingly, they correspond to the trailing end of the scroll wave filament and allow us to analyze its position and velocity.

\begin{figure}
\centering
\includegraphics[width=0.9\linewidth]{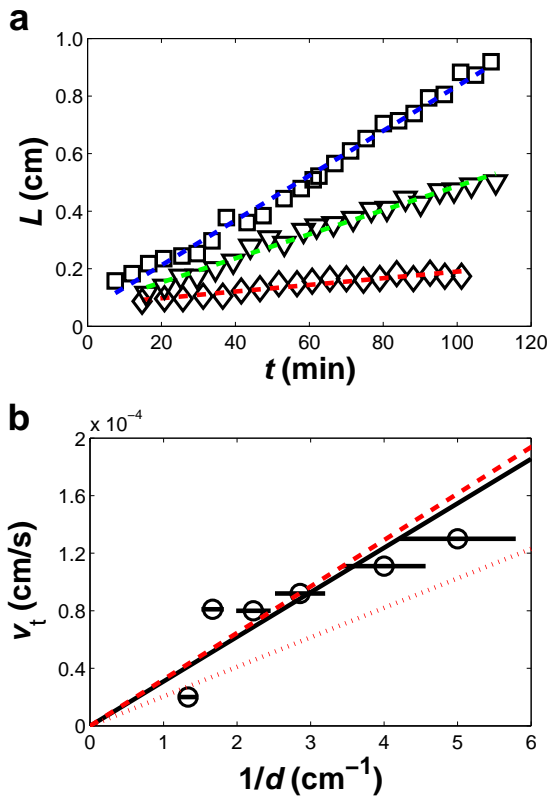}
\caption{\label{fig1} (color online) (a) Temporal evolution of the filament length $L$. Diamonds, triangles, and squares correspond to gap sizes of $d$~= 0.2, 0.45, and 0.75~cm, respectively. The dashed lines are obtained by linear regression of these three data sets. (b) Velocity of the trailing filament terminus as a function of the inverse gap size (open circles). The straight lines are fits assuming $v_t \propto 1/d$ (continuous, black), validity of Eq.~3 (dashed, red), and a circular termination of the filament (dotted, red).}
\end{figure}

Figure~3 analyzes the elongation of partially pinned filaments in more detail. Figure~3(a) shows the temporal evolution of the distance $L$ between the rod and the trailing filament end for three representative experiments that differ only in the gap height $d$. Notice that $L$ is the length of the filament's projection into the image plane. The data sets reveal a linear increase of $L$. The rate of filament elongation equals the difference $v_r-v_t$ between the externally controlled rod speed $v_r$ and the reaction-diffusion-controlled velocity of the trailing filament end $v_t$. Figure~3(b) shows the latter speed as a function of the inverse gap distance $1/d$. In these experiments, the rod speed was kept constant at 0.1~mm/min (i.e. 1.67$\times$10$^{-4}$~cm/s). The data are well described by $v_t = \delta/d$ and yield an average of $\delta$~= 3.1$\times$10$^{-5}$~cm$^2$/s (solid black line). The red lines are discussed later. Notice that $v_t$ cannot be larger than $v_r$ and $1/d$ cannot be smaller $1/h$ (here 0.9~cm$^{-1}$).

\begin{figure}
\centering
\includegraphics[width=0.9\linewidth]{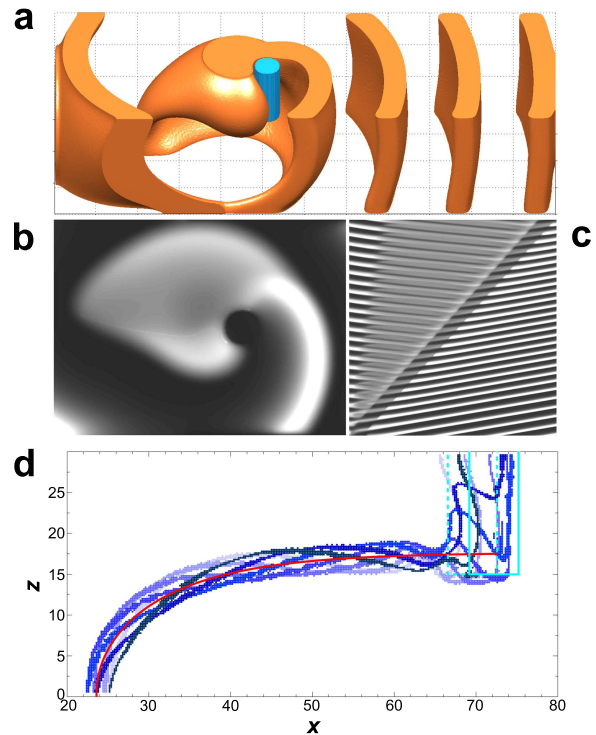}
\caption{\label{fig1} (color online) Numerical simulation of a scroll wave partially pinned to a moving anchor. (a) Snapshot of the three-dimensional wave field $v$ (orange) and the cylindrical heterogeneity (cyan). (b) Partial top view of the same pattern. (c) Time-space plot generated from a sequence of images similar to the one in (b). (d) Superposition of seven filament curves (blueish) obtained during a single rotation period. The anchor moves from the dotted, cyan position to the solid, cyan position. Fitting of Eq.~(2) to the filaments yields the red curve.}
\end{figure}


Our experimental results reveal only two-dimensional projections of the spatially three-dimensional wave patterns and filament shapes. To obtain a better understanding of the unresolved vertical dimension, we performed numerical simulations using the Barkley model \cite{barkley02}:

\begin{align}
   \addtocounter{equation}{1}
   \frac{\partial u}{\partial t} &= D {\nabla}^2 u + {1 \over \epsilon}
       \left\{u(1-u)\left(u-{{v+b}\over{a}}\right)\right\},\tag{2a} \\
\vspace{-0.5cm}
   \frac{\partial v}{\partial t} &= D {\nabla}^2 v + u - v.\tag{2b}
\end{align}

\noindent Although this dimensionless model is not derived from a reaction mechanism, the variables $u$ and $v$ can be associated to the concentrations of the autocatalytic species HBrO$_2$ and ferriin (Fe(phen)$_{3}^{3+}$), respectively. Our simulations use the parameter set $(D, \epsilon, a, b)$~= (1.0, 0.02, 1.1, 0.18) which generates an excitable system in which stable scroll waves exist \cite{Alonso}. Since the diffusion coefficients $D$ in Eqs.~(2a,b) are identical, the filament tension obeys $\alpha = D$ and filaments with small curvature and twist do not move in binormal direction \cite{CGLE97,Alonso}. Accordingly, unperturbed, planar filaments perform curve-shrinking dynamics within their initial plane of confinement. All simulations are based on forward Euler integration with a time step of $6\times10^{-3}$. The box-shaped system is resolved by $600\times200\times150$ grid points at a spacing of $0.2$ and has Neumann boundaries. The moving glass rod is modeled as a translating, cylindrical domain with $(u,v)$~= $(0,0)$. We neglect the Stokes flow generated by the heterogeneity because at the given speed (0.1~mm/min), the fluid motion is noticeable only within a very small region near the spiral center \cite{ke14}. For instance, the typical rod speed and diameter in our experiments cause a creeping flow that decays to about 10~$\%$ over a distance of only 1.2~mm which equals approximately one quarter of the pitch of the free scroll wave.

Figure~4a shows the three-dimensional wave pattern of a vortex that is partially pinned to a rightward moving heterogeneity ($v_r$~= 0.33). This cylinder extends only through the top half of the system. Solid (orange) regions indicate that the local $v$ values are high ($v > 0.2$) and reveal a strongly deformed scroll wave with a rotation backbone that extends from the vortex anchor leftwards. The initial condition of this simulation was an untwisted vortex with a straight, vertical filament and a cylinder placement that matched its horizontal coordinates. The temporal evolution of the wave pattern and the associated filament dynamics show clearly that the filament remains pinned to the moving anchor and that it increases its length at a constant speed. Furthermore, we find that its lower terminus moves rightwards at a speed lower than the speed of the heterogeneity.

Our simulations allow us to generate two-dimensional projections that can be directly compared to our experimental data. For this purpose, we average $v$ over the entire range of vertical $z$ values for each $(x,y)$ location. A representative example of the resulting image data is shown in Fig.~4b. The snapshot qualitatively agrees with the experimental data shown in Fig.~1e. The small differences between our computational and experimental results are likely due to a more pronounced twist of the simulated vortex and/or local effects caused by the Stokes flow in our experiments. Figure~4c is a space-time plot generated from the temporal changes of the projection data. Its overall structure is very similar to the experimental results in Fig.~2b, thus supporting our earlier interpretation.

In the following, we discuss the physical origins of the observed filament dynamics. Figure~4d combines seven snapshots of the filament obtained during one rotation period of the vortex. The overall pattern resembles a bundle of helices. This structure is the result of the local rotation around nearly circular trajectory and some weak twist caused by the partial pinning to the translating anchor. The bundle clearly reveals the stretched out structure of the filament and shows a sharp, nearly perpendicular transition between a horizontal mid-section and the pinned top portion. At the lower terminus, the filament is oriented perpendicular to the system boundary and highly curved. This curvature controls the motion of the trailing end point according to Eq.~(1). We find that the shape of the filament is well described by an analytical solution of Eq.~(1) that had been previously considered in the context of freely moving filaments \cite{dutta10} and ideal grain boundary motion in two dimensions \cite{mullins56}

\begin{equation}
    x(z)=-\frac{\alpha}{v_t}~\ln \cos \bigl( \frac{v_t}{\alpha}(z-z_0) \bigr) + x_0.
\end{equation}

\noindent This curve has a constant hairpin-like shape and moves with a constant speed $v_t$ that is related to the asymptotic, maximal height $w$ of the curve according to

\begin{equation}
v_t = \pi \alpha / (2 w).
\end{equation}

The solid (red) curve in Fig.~4d is the best fit of Eq.~(3) to the helix bundle. Notice that we only evaluate data with $x <$~62 because the abrupt transition to the cylindrical anchor is not captured by this description. We find that the fit captures the shape of the filament bundle well and the asymptotic height ($w$~= 17.3) of the curve is only slightly larger than the gap ($d$~= 15) between the anchor and the lower system boundary. Furthermore, the fit yields $\alpha$~= 1.02 which is very close to the system's known filament tension of 1.0. We conclude that Eq.~(3) provides a very good description of the shape of the elongating filaments.

Equations~(3) and (4) can also be used to interpret our experimental measurements of $v_t$ if we assume that $w = d$. We first establish the filament tension $\alpha$ from independent experiments in which we follow the free collapse of scroll rings. In accordance with Eq.~(1), the radius $R$ of their circular filament obeys $\textrm{d}R/\textrm{d}t = - \alpha / R$ and yields $\alpha$~= 2.05$\times$10$^{-5}$~cm$^2$/s. On the basis of Eq.~(4), this value is used to plot the dashed, red curves in Fig.~3(b). The graph is nearly identical with the proportionality fit (black curve) and hence a good description of the experimental data. For comparison, we also graphed the dependence expected for a trailing filament that terminates as a circular segment of radius $d$. The speed of such as termination point is given by $v_t = \alpha/d$. The slope of the corresponding curve (red dotted line in Fig.~3[b]) is $\pi/2$ times smaller than the slope predicted by Eq.~(4) and does not agree with the experimental results. Lastly we note that Eqs.~(3) and (4) are applicable only to sufficiently elongated filaments as otherwise our approximation of $w = d$ fails.

In conclusion, we have shown that scroll waves can be pinned to moving heterogeneities. Partial pinning of a scroll wave stretches the filament along the trajectory of the anchor. In this process the terminus of the filament is not stationary but follows the anchor at a lower speed that is determined by the filament's local curvature at the system boundary. We expect that filaments can be also stretched out along nonlinear trajectories, which provides a powerful tool for preparing arbitrary shapes including examples that reveal filament interaction and reconnection events \cite{hauser13}.

This material is based upon work supported by the National Science Foundation under Grant No. 1213259.

\end{document}